# Sub-Nyquist sampling boosts targeted light transport through opaque scattering media

Yuecheng Shen[†], Yan Liu[†], Cheng Ma[‡], and Lihong V. Wang[*]

*Optical Imaging Laboratory, Department of Biomedical Engineering, Washington University in St. Louis, One Brookings Drive, St. Louis, Missouri, USA, 63130*
*[†]These authors contribute equally to this work.*
*[‡]Present address: Tsinghua University, Department of Electronic Engineering, Rohm Building 4-108, Beijing 100084, China.*
*[*]Corresponding author: lhwang@wustl.edu*

**Optical time-reversal techniques are being actively developed to focus light through or inside opaque scattering media. When applied to biological tissue, these techniques promise to revolutionize biophotonics by enabling deep-tissue non-invasive optical imaging, optogenetics, optical tweezers and photodynamic therapy. In all previous optical time-reversal experiments, the scattered light field was well-sampled during wavefront measurement and wavefront reconstruction, following the Nyquist sampling criterion. Here, we overturn this conventional practice by demonstrating that even when the scattered field is under-sampled, light can still be focused through or inside opaque media. Even more surprisingly, we show both theoretically and experimentally that the focus achieved by under-sampling is usually about one order of magnitude brighter than that achieved by conventional well-sampling conditions. Moreover, sub-Nyquist sampling improves the signal-to-noise ratio and the collection efficiency of the scattered light. We anticipate that this newly explored under-sampling scheme will transform the understanding of optical time reversal and boost the performance of optical imaging, manipulation, and communication through opaque scattering media. © 2016 Optical Society of America**

*OCIS codes: (110.0113) Imaging through turbid media; (170.7050) Turbid media; (110.1080) Active or adaptive optics; (070.5040) Phase conjugation; (090.2880) Holographic interferometry.*

http://dx.doi.org/10.1364/optica.99.099999

The scattering of light induced by microscopic refractive index inhomogeneity has been a major obstacle to focusing light through or inside opaque scattering media such as biological tissue and fog[1-3]. To overcome this challenge and achieve deep-tissue non-invasive optical imaging, manipulation, therapy, and optical communication through fog and cloud[4-11], various techniques to shape the wavefront of the incident light, including stepwise wavefront shaping[12-17], transmission matrix inversion[18, 19], and optical time-reversal/phase conjugation[20-30], are being actively developed. Among all these techniques, optical time-reversal is most promising for *in vivo* applications, because it achieves the shortest average mode time[31] (the average operation time per degree of freedom) by determining the optimum wavefront globally instead of stepwise.

Optical-time-reversal–based techniques focus light through or inside scattering media by phase conjugating scattered light emitted from a guide star[23, 24, 26, 27, 31-36]. To achieve optical phase conjugation, two types of phase conjugate mirrors (PCMs) have been developed. Analog PCMs employ nonlinear optics based static holography, four-wave mixing, or stimulated Brillouin scattering to generate the phase conjugated field[37, 38]; digital PCMs first employ a digital camera to measure the wavefront of the scattered light with digital holography, and then use a spatial light modulator (SLM) to reconstruct the conjugate wavefront of the scattered light[22, 23]. The pixels of the camera and the SLM are usually one-to-one matched. Although analog PCMs can be fast[39], digital PCMs achieve a much higher phase conjugate reflectivity and have the capability of synthesizing the electric field[28, 33-35], thus becoming more useful and powerful. However, the pixel sizes (several microns to tens of microns) of the digital cameras and SLMs constituting digital PCMs are ~20 times larger than the wavelength of light, so speckle grains are under-sampled (sub-Nyquist sampled, speckle size is half the wavelength) if a PCM is placed adjacent to the rear surface of a thick scattering medium to collect more scattered light from the sample (Fig. 1a, b). Since the Nyquist sampling criterion[40] is not followed, the phase map of the measured wavefront (Fig. 1b) looks different from the real one in Fig. 1a. Such an under-sampled wavefront cannot be exactly time reversed, because it is not a proper representation of the true wavefront. Therefore, in all previous time-reversal–based optical focusing experiments, speckle grains are magnified (Fig. 1c) and are well-sampled by PCM pixels (Fig. 1d), to correctly measure the wavefront of the scattered light[26, 41-43]. The measured wavefront is then used for phase conjugation.

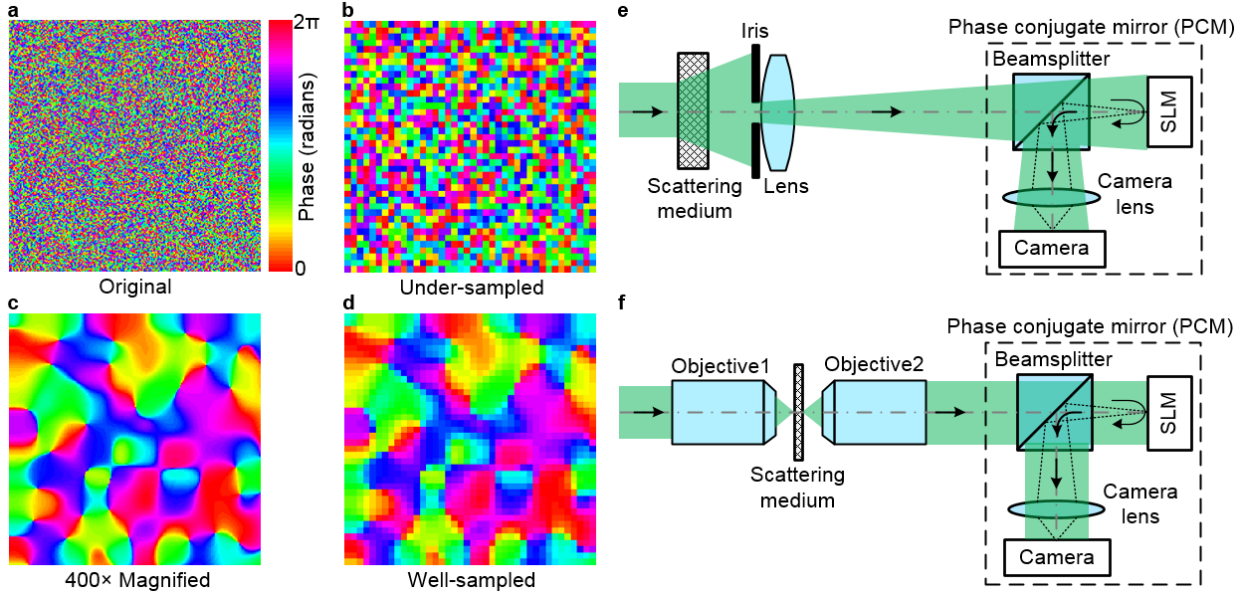

Fig. 1 Sampling of speckle grains in optical time-reversal experiments. a. Phase map of the scattered light on the rear surface of a scattering medium. b. Phase map of the under-sampled speckle grains, which looks different from the real one in a. c. Phase map of the 400× magnified speckle grains. d. Phase map of the well-sampled speckle grains. e. Magnifying the speckle size by an iris and a lens. f. Magnifying the speckle size by two objective lenses.

However, two problems exist with well-sampling speckle grains in time-reversal–based optical focusing experiments. First, the SLM pixels are not efficiently used, since multiple pixels are correlated to represent a single speckle grain. When speckle grains are well-sampled, it has been proved[26] that the peak-to-background ratio (PBR) of the focus (with one speckle inside) is expected to follow

$$\text{PBR} = N_S, \quad (1)$$

where $N_S$ is the number of speckle grains intercepted by the PCM. The PBR is used to quantify the contrast of the focus and is defined as the ratio between the peak intensity of the focus and the average intensity of the speckles when a random wavefront is applied (background). Given that 3×3 pixels to 5×5 pixels have usually been used to sample one speckle grain in previous experiments[26, 41-43], $N_S$ is usually 9 – 25 times smaller than the pixel count of an SLM ($N_P$). Ideally, the PBR should be increased to $N_p$, so that all the degrees of freedom of an SLM can be utilized. Second, to ensure well-sampling speckle grains by magnifying the speckle size, a lens with an iris[44] (Fig. 1e), or two high-magnification objective lenses[23, 25, 29] (Fig. 1f), are usually employed, and the PCM is always placed far from the rear surface of the scattering medium. Consequently, these two approaches detect only a tiny portion of the scattered light exiting the sample and thus have much reduced light collection efficiencies. Moreover, magnifying the speckle size by focusing light onto a scattering medium with an objective (Fig. 1f) works only for thin samples.

In this work, we theoretically predict and experimentally verify that by under-sampling speckle grains, we can not only focus light through scattering media, but also significantly increase the PBR by 9 – 25 times. This discovery overturns the conventional belief that well-sampling speckle grains is required to achieve time-reversal–based optical focusing[26, 41-43]. Besides improving the PBR, since our method does not require magnification of speckle grains, we remove the need to use an iris or objective lenses in the set-up and improve the light collection efficiency. We also proved that under-sampling speckle grains improves the signal-to-noise ratio (SNR) by at least 3 times.

First, by analyzing the time-reversal process, we present a relatively simple argument to show that the expected PBR can be increased to the SLM pixel count $N_P$ when speckle grains are under-sampled in optical time-reversal experiments. A rigorous proof can be found in Supplement 1, Notes 1 and 2. The incident light field $\mathbf{E_{in}}$ is represented by a vector with $N_I$ elements, whose first element is set to 1, and the rest of the elements are set to 0 for simplicity without losing generality. The scattering medium is described by a transmission matrix $\mathbf{T}$ with dimensions of $N_S \times N_I$, whose element $t_{ij}$ is independently drawn from a circular Gaussian distribution[45]. Hence, the scattered light field intercepted by a PCM is computed as[12, 18, 26, 28, 33]

$$\mathbf{E_S} = \mathbf{T}\mathbf{E_{in}} = \begin{pmatrix} t_{11} & t_{12} & \cdots & t_{1N_I} \\ t_{21} & t_{22} & \cdots & t_{2N_I} \\ \vdots & \vdots & \ddots & \vdots \\ t_{N_S 1} & t_{N_S 2} & \cdots & t_{N_S N_I} \end{pmatrix}_{N_S \times N_I} \begin{pmatrix} 1 \\ 0 \\ \vdots \\ 0 \end{pmatrix}_{N_I \times 1} = \begin{pmatrix} t_{11} \\ t_{21} \\ \vdots \\ t_{N_S 1} \end{pmatrix}_{N_S \times 1}. \quad (2)$$

Since speckle grains were well-sampled by a digital camera in previous digital optical phase conjugation experiments, each element of $\mathbf{E_S}$ can be accurately determined by phase-shifting holography[27, 46, 47]. However, when speckle grains are under-sampled, it is unclear what quantity is measured by phase-shifting holography and whether optical focusing using phase conjugation can still be achieved. In Supplement 1, Note 1, we first prove that when speckle grains are under-sampled, the reconstructed quantity of each pixel in phase-shifting holography is the summation of the electric fields of all the speckle grains within that pixel. As an illustration, Figure 2a shows a case where 16 speckle grains are within one pixel. Each speckle grain is assumed to have the same size and shape for simplicity, and the amplitude and the phase of its electric field are represented by the length and the angle of an arrow (the phasor expression). The phase is also encoded by color for better visualization. Using phase-shifting holography, the reconstructed quantity for this pixel equals the summation of the electric fields of all 16 speckle grains (Fig. 2b). We emphasize that it is this field summation, rather than intensity summation, that retains the field information of each speckle grain (although not resolved) and makes optical focusing achievable. With the knowledge that phase-shifting holography measures the summation of the electric fields of all the speckle grains within one pixel, the experimentally measurable scattered field $\mathbf{E}_{S,\text{under-sampled}}$ has the following form when $F$ speckle grains occupy one camera pixel on average:

$$\mathbf{E}_{S,\text{under-sampled}} = \begin{pmatrix} \left.\begin{matrix} t_{11}+t_{21}+\cdots+t_{F1} \\ t_{11}+t_{21}+\cdots+t_{F1} \\ \vdots \\ t_{11}+t_{21}+\cdots+t_{F1} \end{matrix}\right\}F \text{ rows} \\ \vdots \\ \left.\begin{matrix} t_{N_S-F+1,1}+t_{N_S-F+2,1}+\cdots+t_{N_S,1} \\ t_{N_S-F+1,1}+t_{N_S-F+2,1}+\cdots+t_{N_S,1} \\ \vdots \\ t_{N_S-F+1,1}+t_{N_S-F+2,1}+\cdots+t_{N_S,1} \end{matrix}\right\}F \text{ rows} \end{pmatrix}_{N_S\times 1} . \quad (3)$$

By multiplying the backward transmission matrix $\mathbf{T}^T$ (the upper case $T$ stands for matrix transpose) by the conjugated scattered field $\mathbf{E}^*_{S,\text{under-sampled}}$, the optical phase conjugated field $\mathbf{E}_{\mathbf{OPC}}$ exiting the scattering medium can be computed as $\mathbf{E}_{\mathbf{OPC}} = \mathbf{T}^T \mathbf{E}^*_{S,\text{under-sampled}}$. Each element of $\mathbf{E}_{\mathbf{OPC}}$ contains a summation of $F \times N_S$ terms. Among the elements of $\mathbf{E}_{\mathbf{OPC}}$, only the first element (corresponding to the peak) contains a constructive summation of $N_S$ terms $(t_{11} \times t_{11}^* + t_{21} \times t_{21}^* + \cdots + t_{N_S,1} \times t_{N_S,1}^*)$ plus a random summation (random phasor sum) of $(F-1) \times N_S$ terms, while each of the rest of the elements (corresponding to the background) contains a random summation of $F \times N_S$ terms. Thus, the theoretical PBR can be estimated by

$$\text{PBR} = \frac{|E_{\text{peak}}|^2}{|E_{\text{background}}|^2} = \frac{(N_S)^2 + \left(\sqrt{(F-1)\times N_S}\right)^2}{\left(\sqrt{F\times N_S}\right)^2} \approx \frac{N_S}{F} = N_P. \quad (4)$$

We note that the estimated PBR in Eq. (4) is exactly the same as the PBR obtained from a rigorous mathematical derivation (see Supplement 1, Note 2). The above analysis considers using an SLM that achieves full-field (amplitude plus phase) modulation. When other types of SLMs are employed and when speckle grains are under-sampled, we prove that

$$\text{PBR} = \alpha N_P, \quad (5)$$

and $\alpha = \pi/4$, $1/\pi$, and $1/(2\pi)$ for phase-only, binary-phase, and binary-amplitude modulation SLMs, respectively (see Supplement 1, Note 3). The analytical results in Eqs. (4) and (5) are also validated by numerical simulations (see Supplement 1, Note 4). We conclude from the above results that, regardless of the wavefront modulation schemes, light focusing through scattering media can still be achieved even when speckle grains are under-sampled. Moreover, since 3×3 pixels to 5×5 pixels were typically used to sample one speckle grain in previous experiments[26, 41-43], the PBR achieved by under-sampling is 9 – 25 times higher than the PBR achieved by well-sampling, and in this case all the degrees of freedom of an SLM are fully utilized.

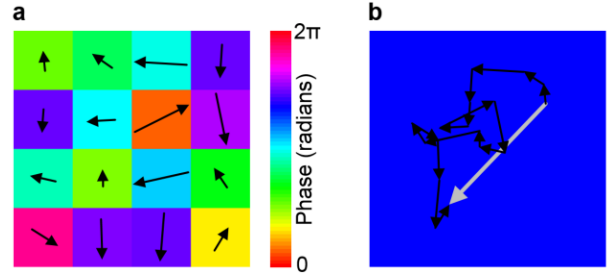

Fig. 2 Physical meaning of the reconstructed quantity in phase-shifting holography when speckle grains are under-sampled. a. An illustration of 16 speckle grains occupying one digital PCM pixel. A phasor expression is used to represent the electric field of each speckle grain. b. The reconstructed quantity (the large gray arrow) is a vector sum of the 16 independent phasors (the small black arrows).

We then performed experiments to investigate time-reversal–based optical focusing through scattering media when speckle grains are under-sampled. The experimental set-up is schematically shown in Fig. 3. The output of a continuous-wave laser (Verdi V10, Coherent) was split into a reference beam (R) and a sample beam (S). Then, each beam was modulated by an acousto-optic modulator (AOM), to induce a $f_b$ =12 Hz frequency difference between R and S. After that, R was expanded by a lens pair, while S was scattered by a scattering medium (SM) composed of three ground glass diffusers (DG-120, Thorlabs). An iris was placed before collecting lens L3 to control the speckle size. S was then combined with R by a beamsplitter (BS), and their interference pattern was recorded by a camera (pco.edge 5.5, PCO-Tech) running at a frame rate of $4f_b$. In this way, we measured the wavefront of S using phase-shifting holography. In the playback process, S was blocked by a mechanical shutter (MS), and the conjugation of the measured phase map was displayed on an SLM (Pluto NIR-II, Holoeye, 1920 × 1080 pixels) whose pixels were one-to-one matched with the camera pixels. Upon reflection off the SLM, R was wavefront shaped and was expected to become a collimated beam after passing through the scattering medium SM. To quantify the quality of time reversal, the light exiting the scattering medium was reflected by a 10:90 BS and focused by lens L4 onto another camera (Camera2, pco.edge 5.5, PCO-Tech).

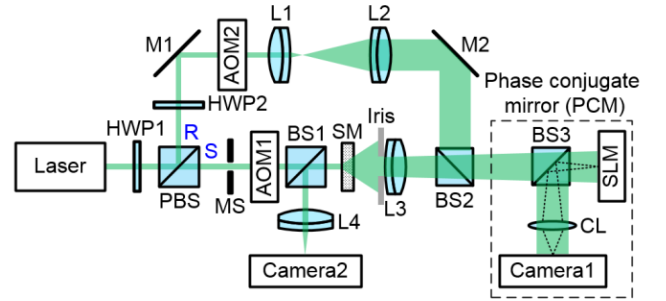

Fig. 3 Schematic of the experimental setup for time-reversal–based optical focusing through scattering media. AOM, acousto-optic modulator; BS, beamsplitter; CL, camera lens; HWP, half-wave plate; L, lens; M, mirror; MS, mechanical shutter; PBS, polarizing beamsplitter; R, reference beam; S, sample beam; SLM, spatial light modulator; SM, scattering medium.

Two experiments were performed to investigate the effect of sampling speckle grains on the quality of time-reversal–based optical focusing through scattering media. In the first experiment, we varied the pixel size of the PCM through pixel binning (the same binning was performed for both Camera1 and the SLM), while fixing the speckle size on the PCM. Without binning, each speckle grain occupied 3.5×3.5 pixels on average. As the pixel size gradually increased by binning pixels, the sampling of speckle grains changed from well-sampled to under-sampled. Figure 5a shows the normalized PBR/$N_P$ as a function of the under-sampling factor $F$. As long as $F$ is no smaller than 1 (speckle grains are under-sampled), PBR/$N_P$ remains close to a constant of 0.117, which is normalized to 1. When $F$ is smaller than 1 (speckle grains are not under-sampled), the normalized PBR/$N_P$ is also smaller than 1, which shows an inefficient utilization of SLM pixels. These experimental results agree with our aforementioned theoretical analysis. In the extreme case when 15×15 pixels are binned to one pixel, corresponding to, on average, 19 speckle grains in one PCM pixel (the far right data point in Fig. 4a), a bright focus with a PBR ~1100 was achieved (Fig. 4b). As a control, when a random phase map was displayed on the SLM, no focus was observed (Fig. 4c).

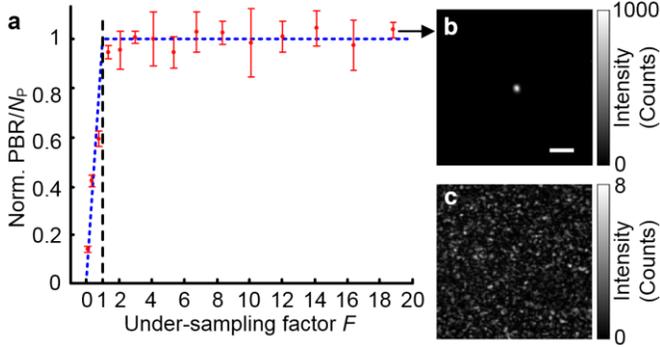

Fig. 4 Experimental results obtained by varying the pixel size of a PCM while fixing the speckle size on the PCM. a. Plot of normalized PBR/$N_P$ as a function of the under-sampling factor $F$. The error bars show the standard deviations obtained from three realizations of the scattering medium. The blue dashed line indicates the theoretical prediction based on Eqs. (1) and (4). b. Image of the achieved focus captured by Camera2 when the under-sampling factor $F$ was 19. The PBR is ~1100. Scale bar, 200 μm. c. No focus was observed when a random phase map was displayed on the SLM.

In the second experiment, we varied the speckle size on the PCM while fixing the pixel size. Super-pixels binned from 5×5 pixels were used throughout this experiment to measure the under-sampling factor when speckle grains were under-sampled. The speckle size was controlled by varying the aperture size of an iris. When the iris was fully opened, a speckle grain occupied ~0.48 super-pixel on average, so it was under-sampled. By gradually closing the iris, the speckle size increased accordingly and finally surpassed the super-pixel size. Figure 6a shows the measured PBR of the focus (normalized by 9100) as a function of the speckle area. When the speckle area is smaller than the super-pixel area (speckle grains are under-sampled), the PBRs are around a constant value of 9100. This observation is consistent with Eq. (5), because the PBR is theoretically determined only by the fixed pixel count and is independent of the under-sampling factor. When the speckle area surpasses the super-pixel area (speckle grains are well-sampled), the PBR is inversely proportional to the speckle area. This observation indicates that the PBR is proportional to $N_S$, which agrees with Eq. (1). Having checked all the PBRs obtained with different speckle areas, we note that higher PBRs were achieved when speckle areas were smaller than the super-pixel area, corresponding to under-sampling speckle grains. Compared with the PBR achieved by using 4.2 × 4.2 super-pixels to well-sample speckle grains, the PBR was improved by 16 times with under-sampling speckle-grains. When the speckle area was 0.48× the super-pixel area (the far left data point in Fig. 5a), a bright focus with a PBR of 9100 was achieved (Fig. 5b). As a control, when a random phase map was displayed on the SLM, no focus was observed (Fig. 5c). Using focused-ultrasound–guided digital optical phase conjugation[24,26-27], we focused light inside a scattering medium comprising two diffusers. Again, the PBR of the focus is higher with under-sampling compared with well-sampling (see Supplement 1, Note 5).

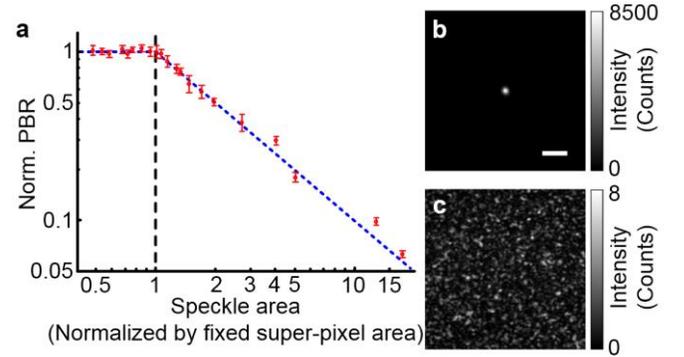

Fig. 5 Experimental results obtained by varying the speckle size on a PCM while fixing the pixel size. a. Plot of normalized PBR as a function of speckle area. The horizontal and vertical axes are shown in the log scale. The blue dashed line indicates the theoretical prediction based on Eqs. (1) and (4). The error bars show the standard deviations obtained from three realizations of the scattering medium. b. Image of the achieved focus when the speckle area was 0.48× the pixel area. The PBR is 9100. Scale bar, 200 μm. c. No focus was observed when a random phase map was displayed on the SLM.

Here, we improved the focusing quality by pushing the upper limit of the theoretical PBR to $N_P$. Although great efforts have been made[41, 43], experimentally achieved PBRs were always lower than their theoretical values, due to misalignment of the system and imperfection of the SLM. In our experiments, when speckle grains were under-sampled and $N_P$ was 83000 (5×5 binning), the achieved PBR was 9100, which is still seven times lower than its theoretical value ($\pi N_P/4 \sim 65000$).

Besides improving the PBR, under-sampling speckle grains also improves the SNR of wavefront measurement. When speckle grains are under-sampled and the main noise source is shot noise, it is proved in Supplement 1, Note 6 that the SNR of wavefront measurement is given by

$$\text{SNR}_{\text{under-sampled}} = 2\sqrt{\text{NPS}}, \quad (6)$$

where NPS is the average number of photoelectrons induced by the light exiting the sample per speckle grain. On the other hand, when speckle grains are well-sampled, it has been proved that the SNR is given by[48]

$$\text{SNR}_{\text{well-sampled}} = 2\sqrt{\text{NPP}} = 2\sqrt{\text{NPS}/G}. \quad (7)$$

Here, NPP is the average number of photoelectrons induced by the light exiting the sample per camera pixel, and $G = 1/F$ describes the average number of pixels used to sample one speckle grain, which is larger than 1. When an iris is used to control the speckle size (Fig. 1e), both the light power and the total number of speckle grains intercepted by the PCM are proportional to the area of the iris aperture, so NPS is a constant, independent of the speckle size. Thus, from Eqs. (6) and (7), we conclude that under-sampling speckle grains increase the SNR of wavefront measurement by a factor of $\sqrt{G}$, compared with well-sampling speckle grains. Considering that 3×3 pixels to 5×5 pixels have usually been used to sample one speckle grain in previous experiments, the SNR can be enhanced by 3 – 5 times with under-sampling speckle grains.

It was proved theoretically that stepwise and time-reversal–based wavefront shaping find the same optimum wavefront[49]. However, before our work, the experimentally determined wavefronts by these two approaches were not the same. Since speckle grains were well-sampled, neighboring pixels were correlated in previous time-reversal–based wavefront shaping, while neighboring (super) pixels were uncorrelated in stepwise wavefront shaping[12]. Under-sampling speckle grains in time-reversal–based wavefront shaping bridges this gap between theoretical prediction and experimental observation, since neighboring pixels are not correlated any longer.

Moreover, we note that when a scattering medium is thick, polarization can be completely scrambled by scattering. By using a vector transmission matrix in the derivation[50,51], it is straightforward to see that all the conclusions we obtained in this work are still valid, except that all the PBRs are reduced by half. Such a PBR reduction can be understood by considering the enhanced background due to the field along a polarization direction orthogonal to the incident polarization direction.

In acoustic time-reversal, Fink *et al.* stated that "the transducers can be spaced as far apart as half the smallest wavelength without impairing the quality of the reproduction"[52,53], suggesting that well-sampling is preferred in acoustic time-reversal. Even though the pitch of a transducer array is two or three times larger than half the acoustic wavelength in some experiments[54,55], it was unclear whether the ultrasonic wavefront was under-sampled in these experiments, since the ultrasonic wave exiting a scattering medium propagated some distance before reaching the array and the ultrasonic coherence area at the array location was not reported. Fink *et al.* later realized that "the best situation would be one in which all array elements receive totally independent information"[56], however, he also stated that "this is not physically possible"[56]. Regardless, to the best of our knowledge, our paper is the first to point out and demonstrate that under-sampling the wavefront is better than well-sampling in optical time-reversal experiments.

In summary, we theoretically and experimentally demonstrate that even when speckle grains are under-sampled, light can still be focused through or inside opaque scattering media. In fact, we proved that sub-Nyquist sampling can boost the PBR by more than ten times than conventional well-sampling conditions and also increase the SNR of wavefront measurement. Moreover, since our method does not require magnification of speckle grains, we remove the need to use an iris or objective lenses in the set-up and are able to place the PCM closer to the sample, thus greatly improving the collection efficiency of the scattered light. We anticipate that our discovery will transform the understanding of optical time-reversal and boost the performance of light focusing through opaque media for optical imaging, manipulation, therapy and communication.

**Funding.** National Institutes of Health (DP1 EB016986, R01 CA186567).

**Acknowledgment.** We thank Prof. James Ballard for proofreading the manuscript.

See Supplement 1 for supporting content.